\documentclass[12pt]{article}
\usepackage{amsmath,amssymb,amsfonts,amsthm}

\title{Dynamics of entropy and nonclassicality features of the interaction between a $\diamondsuit$-type four-level atom and a single-mode field in the presence of intensity-dependent coupling and Kerr nonlinearity}

\author{H R Baghshahi$^{1,2,3}$, M K Tavassoly$^{1,2,*}$ and A. Behjat$^{1,4}$ \\
 \footnotesize{$^1$ Atomic and Molecular Group, Faculty of Physics, Yazd University, Yazd, Iran} \\
 \footnotesize{$^2$ The Laboratory of Quantum Information Processing, Yazd University, Yazd, Iran} \\
 \footnotesize{$^3$ Department of Physics, Faculty of Science, Vali-e-Asr University of Rafsanjan, Rafsanjan, Iran} \\
 \footnotesize{$^4$ Photonics Research Group, Engineering Research Center, Yazd University, Yazd, Iran} \\
 \footnotesize{$^*$ E-mail: mktavassoly@yazd.ac.ir}}

\begin{document}
\maketitle
%=======================================================================
%=======================================
 %=======================================

 \newcommand{\norm}[1]{\left\Vert#1\right\Vert}
 \newcommand{\abs}[1]{\left\vert#1\right\vert}
 \newcommand{\set}[1]{\left\{#1\right\}}
 \newcommand{\R}{\mathbb R}
 \newcommand{\I}{\mathbb{I}}
 \newcommand{\C}{\mathbb C}
 \newcommand{\eps}{\varepsilon}
 \newcommand{\To}{\longrightarrow}
 \newcommand{\BX}{\mathbf{B}(X)}
 \newcommand{\HH}{\mathfrak{H}}
 \newcommand{\A}{\mathcal{A}}
 \newcommand{\D}{\mathcal{D}}
 \newcommand{\N}{\mathcal{N}}
 \newcommand{\x}{\mathcal{x}}
 \newcommand{\p}{\mathcal{p}}
 \newcommand{\la}{\lambda}
 \newcommand{\af}{a^{ }_F}
 \newcommand{\afd}{a^\dag_F}
 \newcommand{\afy}{a^{ }_{F^{-1}}}
 \newcommand{\afdy}{a^\dag_{F^{-1}}}
 \newcommand{\fn}{\phi^{ }_n}
 \newcommand{\HD}{\hat{\mathcal{H}}}
 \newcommand{\HDD}{\mathcal{H}}
%==================================================

 \begin{abstract}
The interaction between a $\diamondsuit$-type four-level atom and a single-mode field in the presence of Kerr medium with intensity-dependent coupling involving multi-photon processes has been studied. Using the generalized (nonlinear) Jaynes-Cummings model, the exact analytical solution of the wave function for the considered system under particular condition, has been obtained when the atom is initially excited to the topmost level and the field is in a coherent state. Some physical properties of the atom-field entangled state such as linear entropy showing the entanglement degree, Mandel parameter, mean photon number and normal squeezing of the resultant state have been calculated. The effects of  Kerr medium, detuning and the intensity-dependent coupling on the temporal behavior of the latter mentioned nonclassical properties have been investigated. It is shown that by appropriately choosing the evolved parameters in the interaction process, each of the above nonclassicality features, which are of special interest in quantum optics as well as quantum information processing, can be revealed.
 \end{abstract}

%==============================================
 %==============================================

 %==========================================================================
 \section{Introduction}
%==========================================================================
In the field of quantum optics, one of the exactly solvable models which describes the interaction between a two-level atom with a single-mode cavity field is the Jaynes-Cummings model (JCM) in the rotating wave approximation (RWA) \cite{Jaynes}. This model has been widely studied both theoretically and experimentally by many researchers during recent decades and many interesting physical features have been remarked upon \cite{Yoo1}. For instance, one may refer to atomic population inversion \cite{Scully,Walls,Perina,Knight,Eberly}, quadrature squeezing \cite{Meystre,Arvinda}, photon anti-bunching \cite{Short}, entanglement \cite{Fang}, atomic dipole squeezing \cite{Li,Naderi} and so on. This model has been extended in different directions such as multi-mode fields \cite{Orany1,Faghihilaser,FaghihiPhysA}, multi-atoms interaction \cite{Mahmood,BaghshahiScr}, multi-level atoms \cite{Cardimona} and Kerr nonlinearity \cite{Joshi}. The interaction between two-level and different types of three-level atoms with quantized cavity field have been reviewed by Yoo and Eberly in \cite{Yoo1}. Also, in recent years, researchers have strongly focused on the nonlinear interaction between a two- or multi-level atom with cavity field which leads to the deformed JCM. The latter nonlinear JCM, which firstly suggested in \cite{Buck,Sukumar} describes the dependence of atom-field coupling on the light intensity. The revival-collapse phenomenon in the quadrature squeezing has been observed in the interaction between a two-level atom with single-mode cavity field of the multi-photon intensity-dependent JCM \cite {Orany2}. A nonlinear Jaynes-Cummings model which constructed from the standard JCM by deforming the single-mode field operators using $f$-deformed oscillator introduced by Man$^{,}$ko et al \cite{Manko} has been studied in \cite{Recamier}.\\
 The nonlinear interaction between a three-level atom in $\Lambda$- and $V$-configuration with a two-mode field under a multi-photon process has been studied respectively in \cite{Obada1} and \cite{Obada2}.
 The interaction between a $\Lambda$- and $V$-type three-level atom with intensity-dependent coupling in a Kerr medium respectively
 studied in \cite{Faghihi} and \cite{Zait1}.
 Recently, one of us has studied the entanglement dynamics of the nonlinear interaction between a $\Lambda$-type three-level
 atom with a two-mode cavity field in the presence of a cross-Kerr medium and its deformed counterpart \cite{Honarasa},
 intensity-dependent atom-field coupling and the detuning parameters \cite{faghihi2,faghihi3}. Also, the authors have investigated
 a three-level atom in motion which interacts with a single-mode field in an optical cavity in an intensity-dependent coupling regime \cite{faghihi4}.
 Sub-Poissonian statistics, population inversion and entropy squeezing of two two-level atoms interacting with a single-mode binomial field have been
 discussed in the intensity-dependent coupling regime in \cite{Hekmat}.
 Very recently, we investigated the effect of Kerr medium, Stark shift, detuning  and time-dependent coupling on the behaviour of quantum
 properties in the full nonlinear interaction regime of a two-level atom with a single-mode cavity field \cite{Baghshahi}.\\
 Besides the above mentioned works, one of the interesting generalizations of the JCM is the interaction between a four-level atom
 with a single-mode cavity field. There are different-types of four-level atoms such as $\lambda$-type, $Y$-type,
 $N$-type, $\diamondsuit$-type, ladder-type and so on. The interaction between different types of four-level atoms
 with a single-mode field has been studied in \cite{Bina,Wang,Gong,Zait2}. The interaction between a four-level $N$-type atom and
 two-mode cavity field in the presence of a Kerr medium has been investigated in \cite{Abdel-Wahab}. The interaction between a
 four-level atom in different configurations (i.e. $Y$, $\lambda$ and ladder configurations) with a single-mode field under
 multi-photon process with additional forms of nonlinearities of both: "field" and "atom-field coupling" has also been studied in
  \cite{Obada3}. In the present work we will deal with the interaction between a $\diamondsuit$-type four-level atom \cite{Morigi}
  with a single-mode cavity field with intensity-dependent coupling and multi-photon processes in a cavity containing a Kerr medium.
   A lot of attentions has been paid to the four-level $\diamondsuit$  atomic configuration in the literature.  For instance, it has
    frequently been demonstrated as a model for observing pressure-induced resonances \cite{Grynberg}. Also, it can be found in experimental
    works with gases of alkali-earth atoms which aim at optical frequency standards \cite{Curtis} or at reaching the quantum degeneracy regime by
    all-optical means \cite{Katori}. In addition, this kind of atomic structure has been proposed for the  generation of ultraviolet laser gain
    through the laser without inversion method \cite{Lukin}. Interestingly, this configuration reduces to $V$- and $\Lambda$-type three-level
    atom by respectively eliminating the topmost  and bottommost  states of the atom. This fact helps us to check our numerical results which
     will be presented. Accordingly, we particularly choose $\diamondsuit$ configuration for the four-level atom in our considered
      matter-wave interaction.\\
%
 %%%%%
One of the main consequences of the above interactions is the appearance of the entanglement.  Entanglement is a key resource which distinguishes quantum information theory from the classical one. It plays a central role in many potential  applications such as quantum information \cite{Nielsen}, quantum teleportation \cite{Benenti}, quantum cryptography \cite{Ye}, quantum computation \cite{Nielsen,Benenti} and dense coding \cite{Ekert}. The degree of entanglement can be evaluated via different measures, such as von Neumann entropy \cite{Plenio}, linear entropy \cite{Kim}, concurrence \cite{Hill} and so on.
 The entropy of the field (atom) is a very useful operational measure of the purity of any atom-field quantum system. The time evolution of the field entropy shows the degree of entanglement (DEM); the higher the entropy, the greater the entanglement.
 In addition, we will pay attention to the most favorite nonclassical features known sub-Poissonian photon statistics \cite{Davidovich} and collapse and revival of atomic population \cite{Scully}.
  Other than the above two criteria, in quantum optics, the squeezing phenomena is described by decreasing the quantum fluctuations in one of the field quadratures with the price of an increase in the corresponding conjugate quadrature. The squeezed light has various applications such as in optical communication networks \cite{Yuen}, gravitational wave detection \cite {Caves} and in optical waveguide tap \cite {Shapiro}.  Furthermore, the generation of squeezed states has been predicted in a number of nonlinear optical systems \cite {Meystre}.
Consequently, in this paper, after obtaining the exact analytical solution of the entire state vector of the system we shall also examine the effects of the Kerr medium, intensity-dependent coupling and detuning on the time evolution of linear entropy, Mandel parameter, mean photon number and normal squeezing. Finally,  the results are compared with the well-known results of  $V$-type and $\Lambda$-type three-level atoms.\\
This paper is organized as follows. In section 2 the generalized Hamiltonian for the atom-field system is introduced and by solving the corresponding Schr\"{o}dinger equation, the probability amplitudes at any time for the whole system has been obtained by specifying particular initial conditions of atom and field. The temporal evolution of the linear entropy (DEM), Mandel parameter, mean photon number,  normal squeezing and the effects of the Kerr medium, detuning and intensity-dependent coupling on the evolution of mentioned properties for single- and two-photon processes are studied in sections 3. At last, a summary and conclusions are presented in section 4.
%
%==================================================================================
 \section{The model and its solution}
 %==================================================================================
 Let us consider a four-level atom with $\diamondsuit$-configuration (has been depicted in Fig. 1) with topmost state $|1\rangle$, ground state $|4\rangle$ and two intermediate states $|2\rangle$ and $|3\rangle$ with the only allowed transitions $|1\rangle\rightarrow|2\rangle$, $|1\rangle\rightarrow|3\rangle$, $|2\rangle\rightarrow|4\rangle$ and $|3\rangle\rightarrow|4\rangle$. Ignoring level $|1\rangle$ and $|4\rangle$   reduces this configuration to three-level $V$- and $\Lambda$-type, respectively. This four-level atom interacts with a single-mode cavity field via an intensity-dependent coupling regime in an optical cavity field in the multi-photon processes involving a Kerr medium.
Based on the JCM formalism, as the full quantum mechanical approach to the problem, our proposed model can be appropriately described by the Hamiltonian:
\begin{eqnarray}\label{1}
\hat{H}=
\sum_{j=1}^{4} \omega_{j}\hat\sigma_{jj}+\Omega \hat{a}^\dagger \hat{a}
+\lambda_{1} (\hat{A}^{k}\hat{\sigma}_{12}+ \hat{A}^{\dagger k}\hat{\sigma}_{21})
+\lambda_{2} (\hat{A}^{k}\hat{\sigma}_{13}+ \hat{A}^{\dagger k}\hat{\sigma}_{31})\nonumber\\\hspace*{-.56in}+\lambda_{3} (\hat{A}^{k}\hat{\sigma}_{24}+ \hat{A}^{\dagger k}\hat{\sigma}_{42})+\lambda_{4} (\hat{A}^{k}\hat{\sigma}_{34}+ \hat{A}^{\dagger k}\hat{\sigma}_{43})\textbf{+}\chi \hat{a}^{\dagger 2} \hat{a}^{2},
\end{eqnarray}
where $\hat{\sigma}_{ij}=|i\rangle \langle j |$ is the atomic raising or lowering operator, $\hat{a}$ and $\hat{a}^\dagger $ are respectively the bosonic annihilation and creation operators of the cavity field, $\chi$ describes the dispersive part of the third-order nonlinearity of the Kerr medium and $\lambda_{j}, j=1,2,3,4$ are the atom-field coupling constants. Also, the deformed operators $\hat{A}$ and $\hat{A}^{\dagger}$ have been defined as:
\begin{equation}\label{2}
\hat{A}=\hat{a} f(\hat{n})= f(\hat{n}+1)\hat{a}, \hspace{1.5cm} \hat{A}^\dagger=f(\hat{n})\hat{a}^\dagger=\hat{a}^\dagger f(\hat{n}+1),
\end{equation}
where $f(\hat{n})$ is a function of the number operator (intensity of light), a well-known operator-valued function in the nonlinear coherent state approach \cite{Manko,Vogel,Roknizadeh}. A deep insight on the Hamiltonian (\ref{1}) indicates that, it is a natural generalization of atom-field interaction with constant coupling by replacing $ \lambda_{j}$ with $\lambda_{j} f(\hat{n})$.\\
In order to obtain the explicit form of the wave function for our system, we solve the time-dependent Schr\"{o}dinger equation $i\frac{\partial}{\partial t}|\psi(t)\rangle=\hat{H}|\psi(t)\rangle$. For the assumed system, the wave function at any time $t$ can be written in the following form:
\begin{eqnarray}\label{3}
|\psi(t)\rangle&=&\sum_{n=0}^{\infty} q_{n}[ A(n,t)e^{-i\gamma_{1}t}|1,n\rangle+B(n+k,t)e^{-i\gamma_{2}t}|2,n+k\rangle\nonumber\\&+& C(n+k,t)e^{-i\gamma_{3}t}|3,n+k\rangle+D(n+2k,t)e^{-i\gamma_{4}t}|4,n+2k\rangle\ ],
 \end{eqnarray}
where $A(n,t)$, $B(n+k,t)$, $C(n+k,t)$ and $D(n+2k,t)$ are the probability amplitudes of finding the atom in states $|1\rangle$, $|2\rangle$, $|3\rangle$ and $|4\rangle$ and
\begin{eqnarray}\label{4}
\gamma_{1}&=&\omega_{1}+\Omega n,
\nonumber\\ \gamma_{2}&=&\omega_{2}+\Omega (n+k),
\nonumber\\ \gamma_{3}&=&\omega_{3}+\Omega (n+k),
\nonumber\\ \gamma_{4}&=&\omega_{4}+\Omega (n+2k).
 \end{eqnarray}
We suppose that the field is initially prepared in the coherent state and the atom enters the cavity in the exited state $|1\rangle$. Thus, the initial wave function is given by:
\begin{equation}\label{5}
|\psi(t=0)\rangle=\sum_{n} q_{n}|1,n\rangle, \hspace{1.5cm}q_{n}=\exp(-\frac{|\alpha|^{2}}{2}) \frac{\alpha^{n}}{\sqrt{n!}}.
 \end{equation}
The equations of motion for the probability amplitudes are obtained by substituting $|\psi(t)\rangle$ from (Eq. \ref{3}) and $\hat{H}$ from (Eq. \ref{1}) in the time-dependent Schr\"{o}dinger equation. Consequently, one arrives at the following four first-order coupled differential equations:
\begin{equation}\label{6}
i\frac{dA(n,t)}{dt}= V_{1}A(n,t)+ f_{1}e^{-i\Delta_{1}t} B(n+k,t)+f_{2}e^{-i\Delta_{2}t} C(n+k,t),
\end{equation}
\begin{equation}\label{7}
i\frac{dB(n+k,t)}{dt}=V_{2}B(n+k,t)+ f_{1}e^{i\Delta_{1}t}A(n,t)+g_{1}e^{-i\Delta_{3}t}D(n+2k,t),
\end{equation}
\begin{equation}\label{8}
i\frac{dC(n+k,t)}{dt}=V_{2}C(n+k,t)+ f_{2}e^{i\Delta_{2}t}A(n,t)+g_{2}e^{-i\Delta_{4}t}D(n+2k,t),
\end{equation}
\begin{equation}\label{9}
i\frac{dD(n+2k,t)}{dt}=V_{3}D(n+2k,t)+g_{1}e^{i\Delta_{3}t}B(n+k,t)+g_{2}e^{i\Delta_{4}t}C(n+k,t),
\end{equation}
where we supposed:
\begin{eqnarray} \label{10}
f_{1}&=& \lambda_{1}\sqrt{\frac{(n+k)!}{n!}}\frac{[f(n+k)]!}{[f(n)]!},\hspace{1.5cm}      f_{2}=\lambda_{2}\sqrt{\frac{(n+k)!}{n!}}\frac{[f(n+k)]!}{[f(n)]!},
\nonumber\\ g_{1}&=& \lambda_{3}\sqrt{\frac{(n+2k)!}{(n+k)!}}\frac{[f(n+2k)]!}{[f(n+k)]!}, \hspace{1.10cm}g_{2}=\lambda_{4}\sqrt{\frac{(n+2k)!}{(n+k)!}}\frac{[f(n+2k)]!}{[f(n+k)]!}, \nonumber\\
V_{1}&=&\chi n(n-1), \hspace{3.90 cm}  V_{2}=\chi (n+k)(n+k-1), \nonumber \\
V_{3}&=&\chi(n+2k)(n+2k-1), \nonumber \\
 \Delta_{1}&=&\omega_{2}-\omega_{1}+k\Omega,\hspace{3.33cm}     \Delta_{2}=\omega_{3}-\omega_{1}+k\Omega, \nonumber\\
 \Delta_{3}&=&\omega_{4}-\omega_{2}+k\Omega,\hspace{3.33cm}     \Delta_{4}=\omega_{4}-\omega_{3}+k\Omega,
 \end{eqnarray}
and  $\left[ f(n)\right]! \doteq f(n)f(n-1) \cdots f(1)$ with $\left[ f(0)\right] !\doteq1$.
Now, by considering the particular condition $\omega_{2}=\omega_{3}$,  the levels $|2\rangle$ and  $|3\rangle$ are indeed degenerate and so $\Delta_{1}=\Delta_{2}$ and $\Delta_{3}=\Delta_{4}$ (see equation (\ref{10})). Also, without loss of generality, one may suppose that, $\lambda_1=\lambda_2=\lambda_3=\lambda_4=\lambda$ which readily leads  one to $f_{1}=f_{2}=f$ and $g_{1}=g_{2}=g$. Under these conditions, the probability amplitudes $B(n+k,t)$ and $C(n+k,t)$ satisfy similar differential equations which lead us to conclude that $B(n+k,t)=C(n+k,t)$. Consequently, the considered atom-field system has appropriate analytical solution.  It ought to be mentioned that, $C(n+k,t)=B(n+k,t)$ has different consequences from the cases $B(n+k,t)=0$ ($C(n+k,t)\neq0$) or $C(n+k,t)=0$ ($B(n+k)\neq0$), the latter cases reduce the $\diamondsuit$-type four-level to $\Xi$-type three-level atoms.
Therefore, the coupled system of differential equations for the probability amplitudes (Eqs. (\ref{6}-\ref{9})) can be reduced to
\begin{equation}\label{11}
i\frac{dA(n,t)}{dt}=V_{1}A(n,t)+2f e^{-i\Delta_{1} t}B(n+k,t),
\end{equation}
\begin{equation}\label{12}
i\frac{dB(n+k,t)}{dt}=V_{2}B(n+k,t)+ f e^{i\Delta_{1} t}A(n,t)+g e^{-i\Delta_{3} t}D(+2k,t),
\end{equation}
 \begin{equation}\label{13}
i\frac{dD(n+2k,t)}{dt}=V_{3}D(n+2k,t)+2ge^{i\Delta_{3} t}B(n+k,t),
\end{equation}
Moreover, it is worth to mention that, even though we are handling  with only three differential equations in our calculations, however, the fourth one still exists as
$i\frac{dC(n+k,t)}{dt}=V_{2}C(n+k,t)+ f e^{i\Delta_{1} t} A(n,t)+g e^{-i\Delta_{3} t} D(n+2k,t)$  and therefore, one can not conclude that the four-level atom has been reduced to three-level system. Our particular condition only leads us to the result $B(n+k, t)=C(n+k, t)$.
By assuming $D(n+2k,t)=e^{i\mu t}$, the Eqs.  (\ref {11}), (\ref{12}) and (\ref{13}) lead us to the third-order algebraic equation
 \begin{equation}\label{14}
\mu^{3}+x_{1}\mu^{2}+x_{2}\mu+x_{3}=0,
\end{equation}
where
\begin{eqnarray} \label{15}
x_{1}&=&V_{1}+V_{2}+V_{3}-\Delta_{1}-2\Delta_{3},
\nonumber\\x_{2}&=&-2(f^{2}+g^{2})+V_{1}V_{2}+V_{1}V_{3}+V_{2}V_{3}-\Delta_{3}(V_{1}+V_{3})-(\Delta_{1}+\Delta_{3})(V_{2}+V_{3}-\Delta_{3}),
\nonumber\\x_{3}&=&-2V_{3}f^{2}+(\Delta_{1}+\Delta_{3}-V_{1})(2g^{2}+V_{3}\Delta_{3}-V_{3}V_{2}).
    \end{eqnarray}
Eq. (\ref{14}) has generally three different roots which may be found \cite{Kardan} as follows:
\begin{eqnarray}\label{16}
\mu_{j}&=&-\frac{1}{3}x_{1}+\frac{2}{3}\sqrt{x_{1}^{2}-3x_{2}}\cos \left(\phi+\frac{2}{3}(j-1)\pi\right),\hspace*{.3in}j=1,2,3,
\nonumber\\ \phi&=&\frac{1}{3}\cos^{-1}\left[ \frac{9 x_{1}x_{2}-2x_{1}^{3}-27x_{3}}{2(x_{1}^{2}-3x_{2})^{3/2}}\right].
\end{eqnarray}
By considering $D(n+2k,t)$ as a linear combination of $ e^{i \mu_{j} t}$ and after straightforward calculations, we obtain the probability amplitudes in the form
\begin{eqnarray}\label{17}
A(n,t)&=& \frac{1}{2f}\sum_{j=1}^{3}[(\mu_{j}+V_{3})(\mu_{j}+V_{2}-\Delta_{3})-2g^{2}] b_{j}\exp[i(\mu_{j}-\Delta_{1}-\Delta_{3})t], \nonumber \\
B(n+k,t)&=&C(n+k,t)= -\frac{1}{2}\sum_{j=1}^{3}(\mu_{j}+V_{3})b_{j}\exp[i(\mu_{j}-\Delta_{3})t], \nonumber \\
 D(n+2k,t)&=&g \sum_{j=1}^{3} b_{j}\exp(i\mu_{j} t),
\end{eqnarray}
where
\begin{eqnarray}\label{18}
b_{j}&=&\frac{2 f}{\mu_{jk}\mu_{jl}},\hspace*{.3in}j \neq k\neq l=1,2,3,
\end{eqnarray}
with $\mu_{jk}=\mu_{j}-\mu_{k}$. The above coefficients for the probability amplitudes satisfy the initial conditions for atom and field ($B(n+k,0)=C(n+k,0)=D(n+2k,0)=0, A(n,0)=1$), recalling that the initial state for the atom is the topmost state $|1\rangle$. Therefore, even though the chosen interacting atom-field system seems to be complicated, we could still find the exact analytical solution of the problem.
By setting $f(n)=1$ in all above relations in this section, one readily obtains the special case of the considered atom-field interaction with constant coupling containing its final solution.

 \section{Physical properties}
Now, which we obtained the probability amplitudes (and so the explicit from of the wave function of the entire system), we are able to study the quantum dynamical properties of the atom and field such as linear entropy, Mandel parameter, mean photon number and normal squeezing. It is worth mentioning that choosing different nonlinearity functions $f(n)$ leads to different physical results. Altogether, in the continuation of this paper,  we select particularly the intensity-dependent coupling as $f(n)=1/ \sqrt{n}$. This function has been obtained by Man$^{,}$ko et al  \cite{Manko}  (where the corresponding coherent states have been named by Sudarshan as harmonious states \cite{Sudarshan}). We particularly choose this function since, in addition, it enables us to compare our further numerical results with $V$- and $\Lambda$-type three-level atoms respectively in Refs. \cite{Zait1} and \cite{Faghihi}.

\subsection{Linear entropy}
As is well-known, entanglement \cite{Fang} is a property that, can be naturally found in the composite quantum systems and perhaps is the most striking feature of quantum optics. The quantum dynamics associated with the presented atom-field quantum system leads to the entanglement between the atom and field. The time evolution of the entropy of the field or atom shows the degree of entanglement (DEM). Among different measures of DEM, in the considered quantum system, we use the linear entropy \cite{Liao} which is defined as:
\begin{equation}\label{21}
S(t)=1-Tr(\rho_{A}^{2}(t)).
\end{equation}
The density matrix of the atom-field system is $\rho_{AF}(t)=|\psi(t)\rangle  \langle \psi(t) |$, where $|\psi(t)\rangle $ is given by Eq. (\ref{3}). Then, the reduced  atomic density matrix can be obtained by tracing over the field as follows
\begin{equation}\label{19}
\rho_{A}(t)=Tr_{F}(\rho_{AF}(t))=\left(
                                   \begin{array}{cccc}
                                     \rho_{11} & \rho_{12} & \rho_{13} & \rho_{14} \\
                                     \rho_{21} & \rho_{22} & \rho_{23} & \rho_{24} \\
                                    \rho_{31} & \rho_{32} & \rho_{33} & \rho_{34} \\
                                     \rho_{41} & \rho_{42} & \rho_{43} & \rho_{44} \\
                                   \end{array}
                                 \right).
\end{equation}
The matrix elements in Eq. (\ref{19}) at any time $t$ can be given as
\begin{eqnarray} \label{20}
\rho_{11}&=&\sum_{n=0}^{\infty}P_{n}A(n,t)A^{*}(n,t),
\nonumber\\ \rho_{22}&=&\sum_{n=0}^{\infty}P_{n}B(n+k,t)B^{*}(n+k,t),
\nonumber\\ \rho_{44}&=&\sum_{n=0}^{\infty}P_{n}D(n+2k,t)D^{*}(n+2k,t),
\nonumber\\ \rho_{12}&=&\sum_{n=0}^{\infty}q_{n+k}q_{n}^{*}A(n+k,t)B^{*}(n+k,t)e^{i\Delta_{1}t},
\nonumber\\ \rho_{14}&=&\sum_{n=0}^{\infty}q_{n+2k}q_{n}^{*}A(n+2k,t)D^{*}(n+2k,t)e^{i(\Delta_{1}+\Delta_{3})t},
\nonumber\\ \rho_{24}&=&\sum_{n=0}^{\infty}q_{n+k}q_{n}^{*}B(n+k,t)D^{*}(n+2k,t)e^{i\Delta_{3}t},
\nonumber\\ \rho_{33}&=&\rho_{32}=\rho_{23}=\rho_{22},\hspace{1.5cm}\rho_{13}=\rho_{21}^{*}=\rho_{31}^{*}=\rho_{12},
\nonumber\\\rho_{41}&=&\rho_{14}^{*},\hspace{3.59cm}\rho_{34}=\rho_{24}^{*}=\rho_{34}^{*}=\rho_{24}.
\end{eqnarray}
Consequently, the linear entropy for the considered atom-field system is in the following form:
\begin{equation}\label{211}
S(t)=1-(\rho_{11}^{2}+4\rho_{22}^{2}+\rho_{44}^{2}+4(|\rho_{12}|^{2}+|\rho_{24}|^{2})+2|\rho_{14}|^{2}).
\end{equation}

In figures 2 and 3, we have plotted the linear entropy of the atom versus scaled time for single-photon (k=1) and two-photon (k=2)  processes for the chosen parameters, respectively. In these figures (and also figures 4-7 which will be presented in the continuation of the paper), left and  right plots concern  respectively with the absence ($f(n)=1$) and presence  ($f(n)=1/\sqrt{n}$) of  the intensity-dependent coupling. Frames (a)  corresponds  to the absence of both Kerr effect and detuning ($\chi=0, \Delta_{1}=\Delta_{3}=0$), (b) corresponds to the presence of Kerr medium and exact resonant condition ($\chi=0.4, \Delta_{1}=\Delta_{3}=0$), (c) deals with the absence of Kerr medium and the nonresonance case ($\chi=0, \Delta_{1}=7, \Delta_{3}=15$)  and finally (d) takes into account the presence of both Kerr medium and detuning ($\chi=0.4, \Delta_{1}=7, \Delta_{3}=15$). As is observed from the figures 2, 3, in general, the maximum values of $S(t)$  are decreased in the intensity-dependent coupling regime, when it is compared with the corresponding figures with constant coupling in all of the cases. In other words, the intensity-dependent coupling causes a reduction in the entanglement degree of the atom-field system. Also, as is seen from the right plot of figure 2(a) (as well as 3(a)), whereas the Kerr medium and detuning are disregarded, a regular oscillatory behavior for the time evolution of the linear entropy is revealed in the presence of intensity-dependent coupling (the same result can be seen in figure 2(c) (as well as 3(c)) where the detuning is entered). Comparing the nonlinear regime of the cases $k=1$ ($k=2$) shows that the presence of both Kerr medium and detuning increases (decreases) the maximum amount of entropy  and so the entanglement degree of the bipartite system. If one compares, in the same manner, the  linear regime of the cases $k=1$ ($k=2$) it may be observed that, depending on the presence or absence of the considered parameters, the entropy can be tuned appropriately. Finally, it should be noticed that, according to our further calculations (not shown here), if we use of the concurrence  for measurement the entanglement between the atom and field by using the following definition \cite{Abdel}
 \begin{equation}\label{22}
C=\sqrt{2 \sum_{i,j=1,2,3,4,i\neq j}(\rho_{ii} \rho_{jj}-\rho_{ij} \rho_{ji})},\hspace{.5cm} \rho_{ij}=\langle i| \rho_{A}(t) | j \rangle, i,j=1,2,3,4
\end{equation}
the obtaining results qualitatively confirm the outcome conclusions from the linear entropy.

\subsection{Quantum statistics}
Among several quantities for the investigation of the nonclassicality of quantum states, we pay attention to the Mandel parameter. This parameter for a single-mode light field has been defined as follows \cite{Mandel}
 \begin{equation}\label{22}
Q=\frac{\langle n^{2}\rangle-\langle n\rangle^2}{\langle n\rangle}-1.
\end{equation}
If $-1\leq Q<0$ ($Q>0$) the field statistics is sub-Poissonian (super-Poissonian) and $Q=0$ shows the Poissonian statistics. Using the wave function for our considered system, it is easily seen that:
\begin{equation}\label{23}
\langle n\rangle=\sum_{n=0}^{\infty}P_{n}\left[ n|A(n,t)|^{2}+2(n+k)|B(n+k,t)|^{2}+(n+2k)|D(n+2k,t)|^{2}\right]
\end{equation}

\begin{equation}\label{24}
\langle n^{2}\rangle=\sum_{n=0}^{\infty}P_{n}\left[ n^{2}|A(n,t)|^{2}+2(n+k)^{2}|B(n+k,t)|^{2}+(n+2k)^{2}|D(n+2k,t)|^{2}\right] ,
\end{equation}
where $P_{n}=|q_{n}|^{2}$ is the Poissonian photon distribution of the initial coherent field and $A(n,t)$, $B(n+k,t)$ and $D(n+2k,t)$ have been determined in Eq. \ref{17}.

We examine the effects of the Kerr medium, intensity-dependent coupling, detuning and multi-photon processes on the temporal evolution of the Mandel $Q$ parameter in figures 4, 5 for single- and two-photon processes, respectively. In the left plots ($f(n)=1$) of figures 4 and 5, Mandel parameter varies between positive and negative values, which means that the photons display super- or sub-Poissonian statistics for different intervals of times, alternatively. The same behavior can be seen in the right plots of figures 4(d), 5(a), 5(b) and 5(d) for the nonlinear case with $f(n)=1/\sqrt{n}$. But, from the right plots of figures 4(a), 4(b) and 4(c), where the intensity-dependent coupling is present, we observe that the Mandel parameter is always negative. Also, the right plot of figure 5(c) shows the full super-Poissonian statistics of field at all times. Comparison between left with right plots of figure 4 which is plotted for $k=1$ (unless the case 4(d))  show that the intensity-dependent coupling disappears the super-Poissonian statistics parts of the field. Typical collapse-revival phenomenon is clearly seen for all of the cases in figures 4, 5, unless the especial cases 4(a), 4(c), 5(a), 5(c) which possess a periodic behaviour. In particular, entering simultaneously  the Kerr and the detuning effects leads to the observation of the collapse and revivals which are a nonclassical feature (compare 4(a) with 4(d) and also 5(a) with 5(d)).

 We end this subsection with an overview on the mean photon number distribution in an explicit manner \cite{Obada3,Singh}. By using Eq. (\ref{23}), in figures 6 and 7, we have depicted the time evolution of mean photon number versus the scaled time $\lambda t$ for the chosen parameters similar to figures 2 and 3, respectively. We can see the collapse-revival phenomena  as a nonclassical sign in all frames of these figures except the right plots of figures 6(a), 6(c), 7(a) and 7(c) which have a regular behaviour in scaled time. In general, comparing 6(a) and 6(d)  indicates that entering the detuning and Kerr effect causes a decrease in the maximum values of mean number of photons (the same situation is observed in 7(a) and 7(d)). In the nonlinear case, when the Kerr medium and detuning are present, the collapse-revival phenomenon is clearly occurred with the single-photon process (see right frames of (b) and (d) in figure 6) as well as with the two-photon process (see right frames of (b) and (d) in figure 7).

\subsection{Normal squeezing}
To investigate the squeezing properties of the field, we introduce two quadrature field operators $\hat{x}=(\hat{a}+\hat{a}^{\dagger})/2$ and $\hat{y}=(\hat{a}-\hat{a}^{\dagger})/2i$.
Therefore, these operators satisfy the uncertainty relation $(\Delta\hat{x})^{2}(\Delta\hat{y})^{2} \geq1/16$. Consequently, a state is said to be squeezed in the variable $\hat{x}$ ($ \hat{y}$) if $(\Delta\hat{x})^{2}<1/4$ ($(\Delta\hat{y})^{2}<1/4$).
Equivalently, if we define $S_{x}=4(\Delta\hat{x})^{2}-1$ and $S_{y}=4(\Delta\hat{y})^{2}-1$
squeezing occurs in $\hat{x}$ ($\hat{y}$) component if $-1<S_{x}<0$ ($-1<S_{y}<0$).
These parameters can be rewritten as follows:
\begin{equation}\label{25}
S_{x} =2 \langle \hat{a}^\dagger \hat{a} \rangle + \langle \hat{a}^{2} \rangle + \langle \hat{a}^{\dagger 2} \rangle - (\langle \hat{a} \rangle + \langle \hat{a}^\dagger \rangle)^{2},
\end{equation}
\begin{equation}\label{26}
S_{y} = 2 \langle \hat{a}^\dagger \hat{a}\rangle - \langle \hat{a}^{2} \rangle - \langle \hat{a}^{\dagger 2} \rangle + (\langle \hat{a}\rangle - \langle \hat{a}^{\dag} \rangle)^{2},
\end{equation}
where $\langle \hat{a}^\dag \hat{a} \rangle$ is given by Eq. (\ref{23}) and the expectation value of arbitrary powers of  the field operator can be easily calculated as:
\begin{eqnarray} \label{27}
\langle\hat{a}^{r}\rangle=\langle\hat{a}^{\dagger r}\rangle^{*} &=& \sum_{n=0}^{\infty}q_{n}^{*}q_{n+r}[\sqrt{\frac{(n+r)!}{n!}}A^{*}(n,t)A(n+r,t)\nonumber\\&+&2 \sqrt{\frac{(n+k+r)!}{(n+k)!}}B^{*}(n+k,t)B(n+k+r,t)\nonumber\\&+&\sqrt{\frac{(n+2k+r)!}{(n+2k)!}}D^{*}(n+2k,t)D(n+2k+r,t)]
\end{eqnarray}
Figure 8 shows the temporal behavior of the normal squeezing in $\hat{x}$-component (left plots) and in $\hat{y}$-component (right plots) in the nonlinear case ($f(n)=1/\sqrt{n}$), without Kerr medium and detuning. Frames (a) and (b) have been plotted for $k=1$ and $k=2$, respectively. We can see from this figure that, the squeezing exists in the $\hat{x}$ quadrature and no squeezing is occurred in the $\hat{y}$ quadrature. Also, with comparing the left plots of 8 one observes that, in addition to the fact that, in the  single-photon process the system is always quadrature squeezed, the depth of squeezing is smaller than the two-photon process. It should be noticed that, according to our further calculations (not shown here), generally, there is no squeezing in the quadratures $\hat{x}$ and $\hat{y}$ for the linear case $f(n)=1$ and also for $f(n)=1/\sqrt{n}$ in the presence of the Kerr medium and/or detuning.
 \section{Summary and conclusion}
 In this paper we studied  the interaction between a degenerate $\diamondsuit$-type four-level atom with a single-mode cavity field, where $k$-photon transition is allowed and the Kerr-medium and intensity-dependent coupling are present. Fortunately, we could solve the dynamical problem and found the explicit form of the wave function of the whole atom-field system analytically, in particular condition where the middle levels of the $\diamondsuit$-type four-level atom are considered to be degenerate. we also  supposed that, the atom enters the cavity in the topmost state and the field is initially prepared in a coherent state. We investigated the linear entropy, Mandel $Q$ parameter, mean photon number and normal squeezing as the most favorite nonclassicality features. Even though our formalism can be used for any nonlinearity function, we particularly have chosen the nonlinearity function $f(n)=1/\sqrt{n}$ for our numerical calculations. The obtained results can be summarized as follows:
\begin{itemize}
  \item The temporal evolution of linear entropy (entanglement), Mandel parameter, mean photon number and normal squeezing are sensitive to Kerr medium and detuning.
      \item The maximum value of the DEM for the linear regime ($f(n)=1$) is greater than the nonlinear regime ($f(n)=1/\sqrt{n}$). Indeed, entering this nonlinearity function reduces the DEM.
      \item The Kerr medium in the presence of detuning decreases the maximum value of DEM, unless for the linear case with two-photon process.
          \item By comparing figures 2(a) and 2(b) with similar figures in  \cite{Zait1} and \cite{Faghihi} , it is clear that, the DEM between atom and field for the $\diamondsuit$-type four-level atom is greater than DEM for the $V$-type and it is smaller than that for the $\Lambda$-type three-level atoms.
              \item The intensity-dependent coupling changes the quantum statistical behaviour of the atom-field state  to a full sub-Poissonian statistics for the single-photon process (unless in  the presence of both Kerr medium and detuning), although its depth of negativity has been reduced. Also, typical collapse-revival, as a nonclassical behaviour is seen.
                  \item The time evolution of the mean photon number shows the collapse-revival phenomenon as a nonclassicality sign of the considered system unless for the nonlinearity case without Kerr effect and detuning (figures 6(a) and 7(a)), and also with detuning (figures 6(c) and 7(c)).
                      \item There is no normal squeezing in the linear case and also nonlinear case in the presence of Kerr medium, detuning and both of them. Moreover, for the nonlinear case in the absence of Kerr effect and detuning it can be observed, too. The depth of the squeezing in the quadrature $\hat{x}$ is greater for two-photon process as compared with the single-photon process.

                           Finally, it ought to be mentioned that, our presented formalism has the potential ability to be applied for all well-known nonlinearity functions such as the center of mass motion of trapped ion \cite{Vogel2}, photon-added coherent states \cite{Agarwal2,Sivakumar}, deformed photon-added coherent states \cite {Safaeian}, $q$-deformed coherent states \cite{Naderi1,Macfalane,Biedenharn,Chaichian} etc as well as different initial atom and radiation field states. But, clearly our presented results are limited to the chosen $f(n)$ for the intensity dependent regime. Obviously, selecting other nonlinearity functions may lead to new and perhaps more interesting conclusions which can be done else where. We have not discussed the effect of the initial field photon number in detail for all cases, but it is obvious that the results may be affected directly by this parameter, as well as all discussed parameters. Altogether, this is also a tunable parameter for achieving the discussed physical features.

\end{itemize}

\section*{Acknowledgements}
The authors would like to thank Dr M J Faghihi for his useful discussions.

%==================================================================

 %==================================================================


\vspace {0.5 cm}

 \begin{thebibliography}{999}


 \bibitem{Jaynes} E. T. Jaynes, F. W. Cummings, Proc. IEEE. \textbf{51} (1963) 89.

\bibitem{Yoo1} H. -I. Yoo, J. H. Eberly, Phys. Rep. \textbf{118} (1985) 239.

 \bibitem{Scully} M. O. Scully, M. S. Zubairy, Quantum Optics, Cambridge University Press, 1997.

 \bibitem{Walls} D. F. Walls, G. J. Milburn, Quantum Optics, Springer, Berlin, 1994.

 \bibitem{Perina} J. Perina, Quantum Statistics of Linear and Nonlinear Optical Phenomena, Reidel, Dordrecht, 1984.

 \bibitem{Knight} P. L. Knight, P. M. Radmore, Phys. Lett. A \textbf{90} (1982) 342.

 \bibitem{Eberly} J. H. Eberly, N. B. Narozhny, J. J. Sanchez-Mondragon, Phys. Rev. Lett. \textbf{44} (1980) 1323.

 \bibitem{Meystre} P. Meystre, M. S. Zubairy, Phys. Lett. A \textbf{89} (1982) 390.

   \bibitem{Arvinda} P. K. Aravind, Hu. Guanghui Hu, Physica B+C \textbf{150} (1988) 427.

  \bibitem{Short} R. Short, L. Mandel, Phys. Rev. Lett. \textbf{51} (1983) 384.

   \bibitem{Fang} M. -F. Fang, S. Swain, P. Zhou, Phys. Rev. A \textbf{63} (2000) 013812.



     \bibitem{Li} X. -S. Li , D. L. Lin, T. F. George, Phys. Rev. A \textbf{40} (1989) 2504.

      \bibitem{Naderi} M. H. Naderi, J. Phys. A: Math. Theor. \textbf{44} (2011) 055304.

 \bibitem{Orany1} F. F. A. El-Orany, J. Phys. A: Math. Gen. \textbf{37} (2004) 6157.

  \bibitem{Faghihilaser} M. J. Faghihi, M. K. Tavassoly, M. Bagheri Harouni, Laser Phys. \textbf{24} (2014) 045202.

  \bibitem{FaghihiPhysA} M. J. Faghihi, M. K. Tavassoly, M. Hatami, Physica A \textbf{407} (2014) 100.

 \bibitem{Mahmood} S. Mahmood, M. S. Zubairy, Phys. Rev. A \textbf{35} (1987) 425.

 \bibitem{BaghshahiScr} H. R. Baghshahi, M. K. Tavassoly, Phys. Scr. \textbf{89} (2014) 075101.


 \bibitem{Cardimona} D. A. Cardimona, Phys. Rev. A \textbf{41} (1990) 5016.

  \bibitem{Joshi} A. Joshi, R. R. Puri, Phys. Rev. A \textbf{45} (1992) 5056.

  \bibitem{Buck} B. Buck, C. V. Sukumar, Phys. Lett. A \textbf{81} (1981) 132.

  \bibitem{Sukumar} C. V. Sukumar, B. Buck, Phys. Lett. A \textbf{83} (1981) 211.

   \bibitem{Orany2} F. F. A. El-Orany, J. Mod. Opt. \textbf{53} (2006) 1699.

   \bibitem{Manko} V. I. Man$^{,}$ko, G. Marmo, E. C. G. Sudarshan, F. Zaccaria, Phys. Scr. \textbf{55} (1997) 528.

   \bibitem{Recamier} O. d. Los. Santos-S\'{a}nchez, J. R\'{e}camier J, J. Phys. B: At. Mol. Opt. Phys. \textbf{45} (2012) 015502.

   \bibitem{Obada1} A. -S. F. Obada, S. A. Hanoura, A. A. Eied, Laser Phys. \textbf{23} (2013) 025201.

    \bibitem{Obada2} A. -S. F. Obada, S. A. Hanoura, A. A. Eied, Laser Phys. \textbf{23} (2013) 055201.

   \bibitem{Faghihi} M. J. Faghihi, M. K. Tavassoly, J. Phys. B: At. Mol. Opt. Phys. \textbf{45} (2012) 035502.

  \bibitem{Zait1} R. A. Zait, Phys. Lett. A \textbf{319} (2003) 461.

       \bibitem{Honarasa} G. R. Honarasa, M. K. Tavassoly, Phys. Scr. \textbf{86} (2012) 035401.

      \bibitem{faghihi2} M. J. Faghihi, M. K. Tavassoly, M. R. Hooshmandasl, J. Opt. Soc. Am. B \textbf{30} (2013) 1109.

      \bibitem{faghihi3} M. J. Faghihi, M. K. Tavassoly, J. Opt. Soc. Am. B \textbf{30} (2013) 2810.

        \bibitem{faghihi4}  M. J. Faghihi, M. K. Tavassoly, J. Phys. B: At. Mol. Opt. Phys. \textbf{46} (2013) 145506.\

        \bibitem{Hekmat} H. Hekmatara, M. K. Tavassoly, Opt. Commun. \textbf{319} (2014) 121.

            \bibitem{Baghshahi} H. R. Baghshahi, M. K. Tavassoly, A. Behjat, Chin. Phys. B \textbf{23} (2014) 074203.


\bibitem{Bina} M. Bina, F. Casagrande, A. Lulli, Eur. Phys. J. D \textbf{49} (2008) 257.

\bibitem{Wang} Y. -H. Wang, L. Hao, X. Zhou, G. L. Long, Opt. Commun. \textbf{281} (2008) 4793.

\bibitem{Gong} K. Wang, Y. Gu, Q-H. Gong, Chin. Phys. \textbf{16} (2007) 130.

\bibitem{Zait2} R. A. Zait, Opt. Commun. \textbf{247} (2005) 367.

\bibitem{Abdel-Wahab} N. H. Abdel-Wahab, J. Phys .B: At. Mol. Opt. Phys. \textbf{40} (2007) 4223.

\bibitem{Obada3} A. A. Eied, S. A. Hanoura, A-S. F. Obada, Int. J. Theor. Phys. \textbf{51} (2012) 2665.

\bibitem{Morigi} G. Morigi, S. Franke-Arnold, G-L. Oppo, Phys. Rev. \textbf{66} (2002) 053409.

\bibitem{Grynberg} G. Grynberg, P. R. Berman, Phys. Rev. A \textbf{41} (1990) 2677.

\bibitem{Curtis} E. A. Curtis, C. W. Oates, L. Hollberg, Phys. Rev. A \textbf{64} (2001) 031403(R).

\bibitem{Katori} H. Katori, T. Ido, Y. Isoya, M. Kuwata-Gonokami, Phys. Rev. Lett. \textbf{82} (1999) 1116.

\bibitem{Lukin} M. D. Lukin, S. F. Yelin, M. Fleischhauer, M. A. Scully, Phys. Rev. A \textbf{60} (1999) 3225.

  \bibitem{Nielsen} M. A. Nielsen, I. L. Chuang, Quantum Computation and Quantum Information, Cambrige University Press, Cambrige, 2000.

   \bibitem{Benenti} G. Benenti, G. Casati, G. Strini, Principles of Quantum Computation and Information, Singapore:World Scientific, 2004.

  \bibitem{Ye} L. Ye, G. -C. Guo, Phys. Rev. A \textbf{71} (2005) 034304.

  \bibitem{Ekert} A. K. Ekert, Phys. Rev. Lett. \textbf{67} (1991) 661.

   \bibitem{Plenio} M. B. Plenio, V. Verdal, Phys. Rev. A \textbf{57} (1998) 1619.

   \bibitem{Kim} J.I. Kim, M.C. Nemes, A.F.R. de Toledo Piza, H.E. Borges, Phys. Rev. Lett. \textbf{77} (1996) 207.

\bibitem{Hill} S. Hill, W.K. Wootters, Phys. Rev. Lett. \textbf{78} (1997) 5022.

  \bibitem{Davidovich} L. Davidovich, Rev. Mod. Phys. \textbf{68} (1996) 127.

\bibitem{Yuen} H. P. Yuen, J. H. Shapiro, IEEE Trans. Inf. Theory \textbf{26} (1980) 78.

\bibitem{Caves} C. M. Caves C M, Phys. Rev. D \textbf{23} (1980) 1693.

\bibitem{Shapiro} J. H. Shapiro, Opt. Lett. \textbf{5} (1980) 351.



 \bibitem{Vogel} R. L. de Matos Filho, W. Vogel, Phys. Rev. A \textbf{54} (1996) 4560.

 \bibitem{Roknizadeh} R. Roknizadeh, M. K. Tavassoly, J. Phys. A: Math. Gen. \textbf{37} (2004) 8111.

\bibitem{Kardan} L. N. Childs, A Concrete Introduction to Higher Algebra 3rd edn, Berlin: Springer, 2008.

  \bibitem{Sudarshan} E. C. G. Sudarshan, Int. J. Theor. Phys. \textbf{32} (1993) 1069.

  \bibitem{Liao} Q. Liao, G. Fang, Y. Wang, M. A. Ahmad, S. Liu, Optik. \textbf{122} (2011) 1392.

    \bibitem{Abdel} M. Abdel-Aty, Laser Phys. \textbf{16} (2006) 1381.

    \bibitem{Mandel} L. Mandel, Opt. Lett. \textbf{4} (1979) 205.
    \bibitem{Singh} S. Singh, C. H. Raymond Ooi, Amrita, Phys. Rev. A \textbf{86} (2012) 023810.

    \bibitem{Vogel2} R. L. de Matos Fillho, W. Vogel, Phys. Rev. A \textbf{58} (1998) R1661.

    \bibitem{Agarwal2} G. S. Agarwa, K. Tara, Phys. Rev. A \textbf{43} (1991) 492.

     \bibitem{Sivakumar} S. Sivakumar, J. Phys. A: Math. Gen. \textbf{32} (1999) 3441.

  \bibitem{Safaeian} O. Safaeian, M. K. Tavassoly, J. Phys. A: Math. Theor. \textbf{44} (2011) 225301.


   \bibitem{Naderi1} M. H. Naderi, M. Soltanolkotabi, R. Roknizadeh, J. Phys. Soc. Jpn. \textbf{73} (2004) 2413.

    \bibitem{Macfalane} A. J. Macfarlane, J. Phys. A: Math. Gen. \textbf{22} (1989) 4581.

   \bibitem{Biedenharn} L. C. Biedenharn, J. Phys. A: Math. Gen. \textbf{22} (1989) L873.

     \bibitem{Chaichian} M. Chaichian, P. Kulish, Phys. Lett. B \textbf{234} (1990) 72.
  \end{thebibliography}
 \end{document}